\documentclass{article}
 \usepackage[dvipdf]{epsfig}
  \usepackage{color}
 \usepackage{epsfig}
\usepackage{color}

\begin{document}
\def\q{\texttt{q}}
\def\be{\begin{equation}}
\def\ee#1{\label{#1}\end{equation}}
\newcommand{\ben}{\begin{eqnarray}}
\newcommand{\een}{\end{eqnarray}}
 
\date{}
\title{ Fermions in a Walecka-type cosmology}
\author{Marlos O. Ribas\footnote{gravitam@yahoo.com},$\,$  Pedro  Zambianchi Jr.\footnote{zambianchi@utfpr.edu.br} \\
Departamento de F\'{\i}sica, Universidade Tecnol\'ogica Federal do Paran\'a,
 Curitiba, Brazil\\
Fernando P. Devecchi\footnote{devecchi@fisica.ufpr.br}$\,$ and  Gilberto M. Kremer\footnote{kremer@fisica.ufpr.br}
  \\ Departamento de F\'{\i}sica,
Universidade Federal do Paran\'a, Curitiba, Brazil
}

\maketitle
\begin{abstract}
A simplified Walecka-type model is investigated in a cosmological scenario.
The model includes  fermionic, scalar and  vector fields as sources. It is
shown that  their interactions, taking place in a Robertson-Walker metric, could be responsible for the transition of accelerated-decelerated periods in the early universe
and a current accelerated regime. It is also discussed the role of the fermionic field as the promoter of the accelerated regimes in the early and the late stages of the universe.
\end{abstract}

\section{Introduction}
One of the most fundamental discoveries in observational  cosmology showed that the universe is not only expanding but it does at an accelerated rate \cite{nature}. As the source of this expanded acceleration remains unknown, several candidates have been proposed to account for it. They include, among others, cosmological constant, matter with exotic equation of state, scalar fields \cite{1}, as an attempt  to describe the early stages of the inflationary universe and current acceleration. Alternatively, fermionic fields \cite {2,3} have been successfully used, leading to earlier accelerated era  as well as later decelerated and accelerated stages.   {These fermionic sources have been investigated by using
several  approaches, including numerical and exact solutions,   perturbations, dark spinors, anisotropy-to-isotropy scenarios
and cyclic cosmologies (see, for example \cite{2,3}).} When considering these models, a key point is the choice of the interaction potentials. In previous works several self-interaction fermionic potentials were tested \cite{2,3,3A,4},  like Nambu-Jona-Lasinio, and Yukawa type  \cite{birrel}. In \cite{4} the  authors developed a model in which a fermionic field interacting through an Yukawa-type potential led to  different dynamical regimes of the universe. Another possibility that replaces dark energy makes use of noninteracting massless vector fields alone to derive inflation \cite{5}.
Our approach to cosmological dynamics is inspired by a relativistic mean field model  known as Walecka model \cite{8}.
  {In the  Walecka model, the relativistic  nucleus   interactions occur via the exchange of virtual  mesons, with the
Lagrangian controlling  these interactions with additional terms \cite{8}. The  nucleons obey the Dirac equation, while the (scalar) virtual  mesons obey the Klein-Gordon equation. Although the theory structure is invoked here, we suggest a completely different interpretation, reinforced by the fact that we are supposing an early universe, leaving the inflationary period. The consequences of these interactions to the fate of the universe evolution is one point of discussion in this work. Hence,}
 our goal is to investigate whether the presence of a massless fermionic, a massive scalar and a massive vector fields interacting in the Robertson-Walker metric may have led to different cosmological regimes. Our model shows that this is indeed the case in that all three regimes - initial acceleration, deceleration and later acceleration - are present. One important consequence is that the fermionic field is the promoter of the accelerated regimes in the early and the late stages of the universe.

\section{The Walecka-type model}

We consider that the sources of the gravitational field are related to:
\begin{enumerate}
  \item [(\textbf{a})] the Lagrangian density of a massless fermionic field with self interaction potential $V(\overline\psi\psi)$
\begin{equation}
\mathcal{L}_{f}=\frac{\imath}{2}[ \overline\psi\,\Gamma^\mu
D_\mu\psi-(D_\mu\overline\psi)\Gamma^\mu\psi]-V(\overline\psi\psi),
\label{1}
\end{equation}
where $\psi$ and $\overline\psi=\psi^\dag\gamma^0$ represent the spinor field and its adjoint, respectively. Due to the principle of general covariance,  the Dirac-Pauli matrices $\gamma^a$ are replaced by their generalized versions $\Gamma^\mu=e^\mu_a\gamma^a$, where $e^\mu_a$ are the tetrad fields, and the $\Gamma^\mu$ matrices satisfy
 the generalized Clifford algebra $\{\Gamma^\mu,\Gamma^\nu\}=2g^{\mu\nu}$. { Following the principle of general covariance, the Latin index
corresponds to  the local Lorentz frame and the Greek one to the general frame. Furthermore, $\gamma^a$ are the 4x4 Pauli-Dirac matrices
 \ben
 \gamma^0=\left(
            \begin{array}{cc}
              1 & 0 \\
              0 & -1 \\
            \end{array}
          \right),\qquad \gamma^i=\left(
                           \begin{array}{cc}
                             0 & \sigma^i \\
                             -\sigma^i & 0 \\
                           \end{array}
                         \right),
 \een
 where $\sigma^i$ are the 2x2 Pauli matrices.}
 The covariant derivatives in (\ref{1}) read
\ben\label{2a}
D_\mu\psi= \partial_\mu\psi-\Omega_\mu\psi+\imath\q A_\mu\psi,\\
D_\mu\overline\psi=\partial_\mu\overline\psi+\overline\psi\Omega_\mu-\imath\q\overline\psi A_\mu.
 \label{2b}
\een
Above $\q$ is a constant which couples the fermionic field with the vector field $A_\mu$. Moreover, $\Omega_\mu$ is  the spin connection
\begin{equation}
\Omega_\mu=-\frac{1}{4}g_{\rho\sigma}[\Gamma^\rho_{\mu\delta}
-e_b^\rho(\partial_\mu e_\delta^b)]\Gamma^\delta\Gamma^\sigma,
\label{3}
\end{equation}
with $\Gamma^\nu_{\sigma\lambda}$ denoting the Christoffel symbols;
  \item [(\textbf{b})] the Lagrangian density of a massive scalar field $\phi$ without self interaction potential
\begin{equation}
\mathcal{L}_{b}=\frac{1}{2}\partial^\mu\phi\partial_\mu\phi-\frac{1}{2}m_{b}^2\phi^2,
\label{4}
\end{equation}
with $m_{b}$ denoting the mass of the scalar field;
  \item [(\textbf{c})] the Lagrangian density of a massive vectorial field  $A_\mu$
  \be
  \mathcal{L}_{v}=\frac{1}{2}m_{v}^2A_\mu A^\mu-\frac{1}{4} F_{\mu\nu}F^{\mu\nu},
  \ee{5}
  where  $m_{v}$ is the mass of the vectorial field and $F_{\mu\nu}=\partial_\mu A_\nu-\partial_\nu A_\mu$;
  \item [(\textbf{d})] the Lagrangian density that corresponds to the Yukawa interaction between  the fermionic and the scalar fields
  \be
  \mathcal{L}_{Y}=-\lambda\overline\psi\phi\psi,
  \ee{6}
  with $\lambda$ representing the coupling constant of the Yukawa potential.
\end{enumerate}
Hence, the  action of the model in its explicit form  reads
\ben\nonumber
S=\int\sqrt{-g}d^4x\Bigg\{\frac{R}{2}+\frac{1}{2}\partial^\mu\phi\partial_\mu\phi-\frac{1}{2}m_{b}^2\phi^2
+\frac{i}{2}\left[\overline\psi\Gamma^\mu D_\mu\psi-(D_\mu\overline\psi)\Gamma^\mu\psi\right]-V(\overline\psi\psi)
\\\label{7}-\lambda\overline\psi\phi\psi+\frac{1}{2}m_{v}^2A_\mu A^\mu-\frac{1}{4}F_{\mu\nu}F^{\mu\nu}\Bigg\}.
\een
In the above action $R$ denotes the curvature scalar and it was consider $8\pi G=1$.

In order to study the evolution of a homogeneous and isotropic spatially flat universe, we use the Robertson-Walker metric
\begin{equation}ds^2=dt^2-a(t)^2(dx^2+dy^2+dz^2),\end{equation}
where $a(t)$ is the cosmic scaling factor. In this metric, the components of the tetrad, Dirac-Pauli matrices and spin connection become
\ben
e^\mu_0=\delta^\mu_0, \qquad e^\mu_i=\frac{1}{a(t)}\delta^\mu_i, \qquad \Gamma^0=\gamma^0,\\
\Gamma^i=\frac{1}{a(t)}\gamma^i, \qquad \Omega_0=0, \qquad\Omega_i=\frac{1}{2}\dot a(t)\gamma^i\gamma^0,
\een
where the dots denote time derivatives.

Here we shall investigate the case where the vector field is time-like, namely,
\begin{equation}
A_\mu=(A_0(t),0,0,0),
\end{equation}
which is the  only possible ansatz consistent with a homogeneous and isotropic universe that leads to a diagonal stress-energy tensor with components  $T_{11}=T_{22}=T_{33}$.
This hypothesis implies that the antisymmetric tensor vanishes, i.e., $F_{\mu\nu}\equiv0$. Furthermore,  we shall adopt  that the self interaction potential of the fermionic field is given by $V(\overline\psi\psi)=\xi\left(\overline\psi\psi\right)^n$, where $\xi$ and $n$ are constants.  The chiral
symmetry breaking term  reduces to a fermionic mass term when $n=2$.

Due to the hypothesis of homogeneity and isotropy the fermionic and scalar fields depend only on time, so that we may obtain through a partial integration of (\ref{7}) the point-like Lagrangian:
\ben\label{8}
{\cal{L}}=3a\dot a^2-a^3\frac{\imath}{2}\left[\overline\psi\gamma^0\dot{\psi}-\dot{\overline\psi}\gamma^0\psi+2\imath\q A_0\overline\psi\gamma^0\psi\right]
+a^3\left[\xi(\overline\psi\psi)^n +\lambda\overline\psi\phi\psi-\frac{1}{2}\dot\phi^2+\frac{1}{2}m_b^2\phi^2-\frac{1}{2}m_v^2A_0^2\right].
 \een

 From the Euler-Lagrange equations for the fields $\psi$ and $\overline\psi$ applied to the point-like Lagrangian (\ref{8})  follows the Dirac equations for the spinor field and its adjoint, namely,
\be
\dot{\psi}+\frac{3}{2}\frac{\dot a}{a}\psi+\imath\left[\q A_0+n\xi(\overline\psi\psi)^{n-1}\gamma^0+\lambda\phi\gamma^0\right]\psi=0,
\ee{9}
\be
\dot{\overline\psi}+\frac{3}{2}\frac{\dot a}{a}\overline\psi-\imath\overline\psi\left[\q A_0+n\xi(\overline\psi\psi)^{n-1}\gamma^0+\lambda\phi\gamma^0\right]=0,
\ee{10}

The Klein-Gordon equation is obtained from the Euler-Lagrange equation for $\phi$ and reads
\be
\ddot {\phi}+3\frac{\dot a}{a}\dot{\phi}+\lambda\overline\psi\psi+m_b^2\phi=0,
\ee{11}

The component $A_0$ of the vector field can be determined from the Euler-Lagrange equation, yielding,
\be
A_0=\q\frac{\overline\psi\gamma^0\psi}{m_v^2},
\ee{12}
which shows that it can be determined once we know the time evolution of $\overline\psi\gamma^0\psi$.

By imposing that the energy function associated with the point-like Lagrangian vanishes, namely,
\be
{\cal{E}}=\frac{\partial{\cal{L}}}{\partial\dot a}\dot a+\frac{\partial{\cal{L}}}{\partial\dot{\phi}}\dot{\phi}+\frac{\partial{\cal{L}}}{\partial\dot {\psi}} \dot{\psi}+\dot{\overline\psi}\frac{\partial{\cal{L}}}{\partial\dot {\overline\psi}}-{\cal{L}}=0.
\ee{13}
it follows the  Friedmann equation
\be
3\left(\frac{\dot a}{a}\right)^2=\rho,
\ee{14}
where   the energy density  of the sources of the gravitational field $\rho$ is given by
\be
\rho=\frac{1}{2}\dot{\phi^2}+\frac{1}{2}m_b^2\phi^2+\xi(\overline\psi\psi)^n+\lambda\overline\psi\phi\psi
+\frac{1}{2}\frac{\q^2}{m_v^2}(\overline\psi\gamma^0\psi)^2.
\ee{15}

Finally, the acceleration equation is obtained from the Euler-Lagrange equation for $a$:
 \be
 \frac{\ddot a}{a}=-\frac{1}{6}\left(\rho+3p\right).
 \ee{16}
In the above equation the pressure of the sources of the gravitational field reads
\be
p=\frac{1}{2}\dot{\phi^2}-\frac{1}{2}m_b^2\phi^2+\xi(n-1)(\overline\psi\psi)^n+\frac{1}{2}\frac{\q^2}{m_v^2}(\overline\psi\gamma^0\psi)^2.
\ee{17}

If we multiply the Dirac equation  (\ref{9}) by $\overline\psi$ and sum with (\ref{10}) multiplied by $\psi$ we get the following equation for the bilinear $\overline\psi\psi$
\be
\dot{\overline\psi}\psi+\overline\psi\dot\psi+3\frac{\dot a}{a}\overline\psi\psi=0,
\ee{18}
which furnishes through integration
\be
\overline\psi\psi=\frac{C}{a^3},\qquad \hbox{where}\qquad C=\hbox{constant}.
\ee{19}
Hence, the bilinear decays like a pressureless matter field.

Instead of using the Friedmann equation (\ref{14}) we shall deal with the conservation law of the energy density, namely,
\be
\dot\rho+3\frac{\dot a}{a}(\rho+p)=0,
\ee{20}
which follows from (\ref{14}) and (\ref{16}).

Now if we  substitute the definitions of the energy density density (\ref{15}) and of the pressure (\ref{17}) into (\ref{20}) and use the Klein-Gordon equation (\ref{11}) we obtain the following differential equation for $\overline\psi\gamma^0\psi$
\be
\dot{\overline\psi}\gamma^0\psi+{\overline\psi}\gamma^0\dot\psi+3\frac{\dot a}{a}\left(\overline\psi\gamma^0\psi\right)=0,
\ee{21}
whose  integration leads to
\be
\left(\overline\psi\gamma^0\psi\right)=\frac{C'}{a^3},\qquad \hbox{where}\qquad C'=\hbox{constant}.
\ee{22}
Hence, we infer from (\ref{12}) and (\ref{22}) that the component $A_0$ of the vector field decays also as a pressureless matter field.

With the results (\ref{19}) and (\ref{22}) we may write  the Klein-Gordon  (\ref{11}) and acceleration  (\ref{16}) equations  as
\ben\label{23}
\ddot\phi+3\frac{\dot a}{a}\dot\phi+\frac{C_1}{a^{3}}+C_2\phi=0,
\\\label{24}
\frac{\ddot a}{a}=-\frac{1}{6}\left[2\dot\phi^2+\frac{C_1\phi}{a^3}-C_2\phi^2+\frac{C_3}{a^6}+\frac{C_4(3n-2)}{a^{3n}}\right],
\een
respectively, where we have introduced new constants
\ben\nonumber
C_1=\lambda C, \quad C_2=m_b^2, \quad C_3=\frac{2\q^2C'^2}{m_v^2}, \quad C_4=\xi C^n.
\een
Furthermore the energy density (\ref{15}) and the pressure (\ref{17}) become
\ben\label{ee1}
\rho=\frac{1}{2}\dot{\phi^2}+\frac{C_1\phi}{a^3}+\frac{C_2\phi^2}{2}+\frac{C_3}{4a^6}+\frac{C_4}{a^{3n}},\\\label{ee2}
p=\frac{1}{2}\dot{\phi^2}-\frac{C_2\phi^2}{2}+\frac{C_3}{4a^6}+(n-1)\frac{C_4}{a^{3n}}.
\een
{We note that the pressure equation (\ref{ee2}) can be decomposed  into partial  pressures of the scalar $p_\phi$, fermionic $p_\psi$  and vector $p_A$ fields as follows:
\be
p_\phi=\frac{1}{2}\dot{\phi^2}-\frac{C_2\phi^2}{2},\quad
p_\psi=(n-1)\frac{C_4}{a^{3n}},\quad
p_A=\frac{C_3}{4a^6}.
\ee{pp}
The above decomposition will be useful for the identification of the positive and negative contributions related to the acceleration field.}

Equations (\ref{23}) and (\ref{24}) represent a system of coupled nonlinear differential equations for the fields $a(t)$ and $\phi(t)$.
A numerical solution of this system of equations can be found by specifying the initial conditions for the cosmic scaling factor $a(0)$ and for the scalar field $\phi(0)$ as well as for their derivatives $\dot a(0)$ and $\dot\phi(0)$.

\begin{figure}\vskip1cm
\begin{center}
\includegraphics[width=7cm]{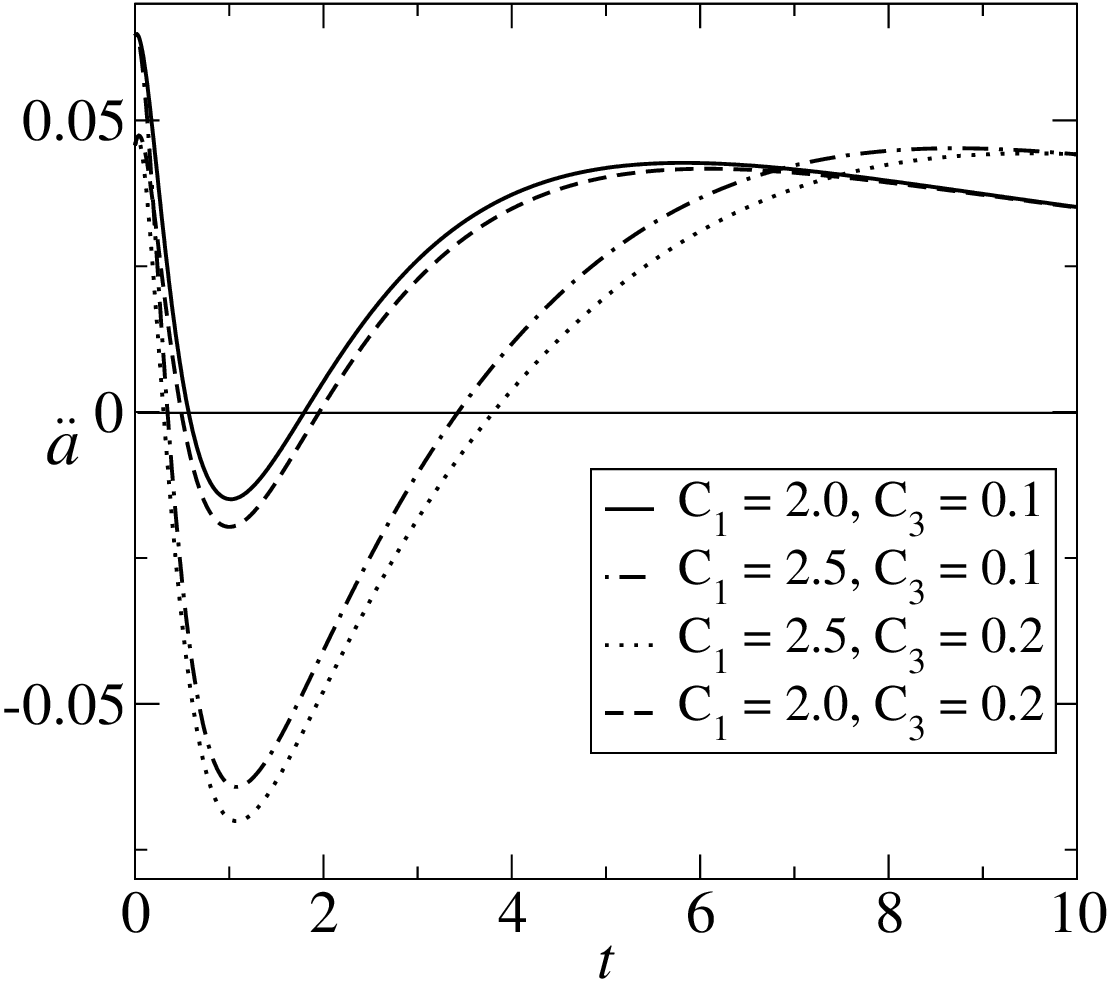}
\caption{Acceleration $\ddot a$ versus time $t$.}
\end{center}
\end{figure}

\begin{figure}\vskip1cm
\begin{center}
\includegraphics[width=7cm]{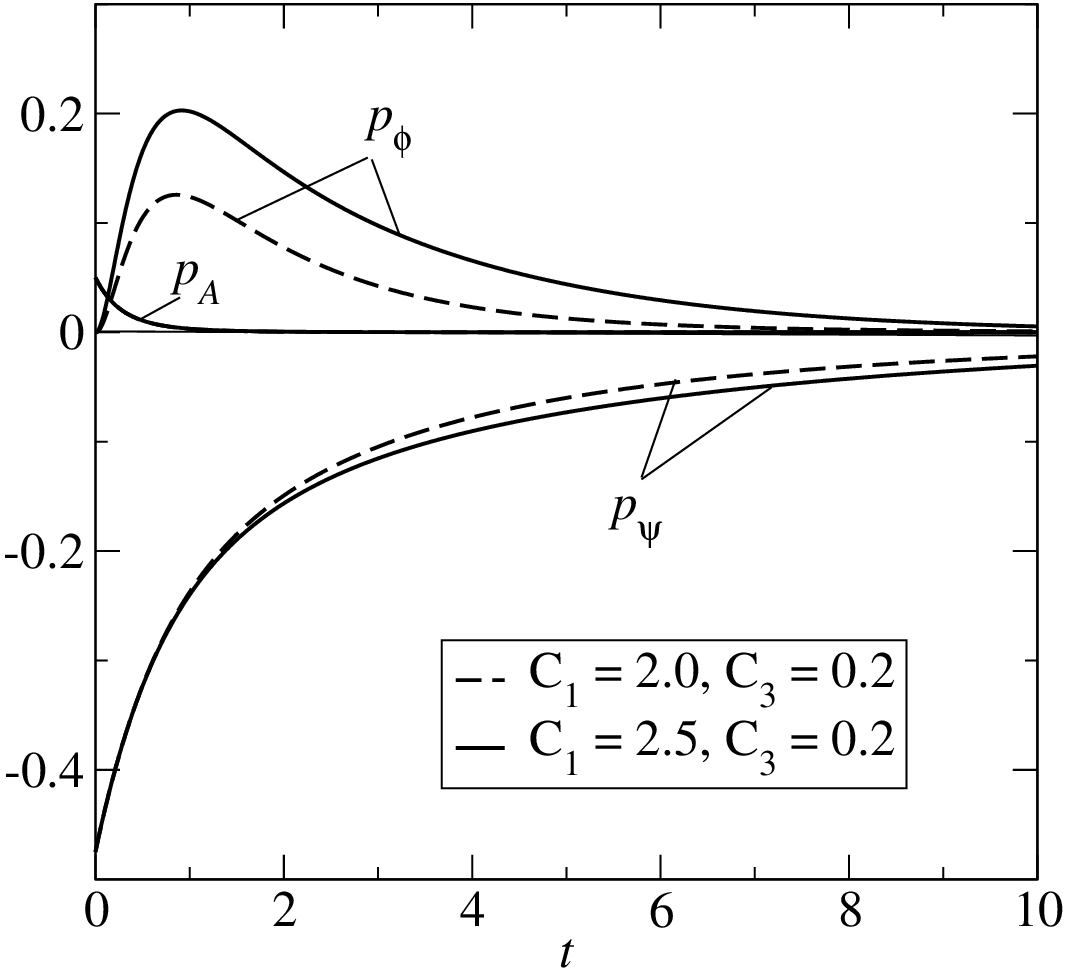}
\caption{Fermionic $p_\psi$, scalar $p_\phi$ and vector $p_A$ pressures versus time $t$. }
\end{center}
\end{figure}

\begin{figure}\vskip1cm
\begin{center}
\includegraphics[width=7cm]{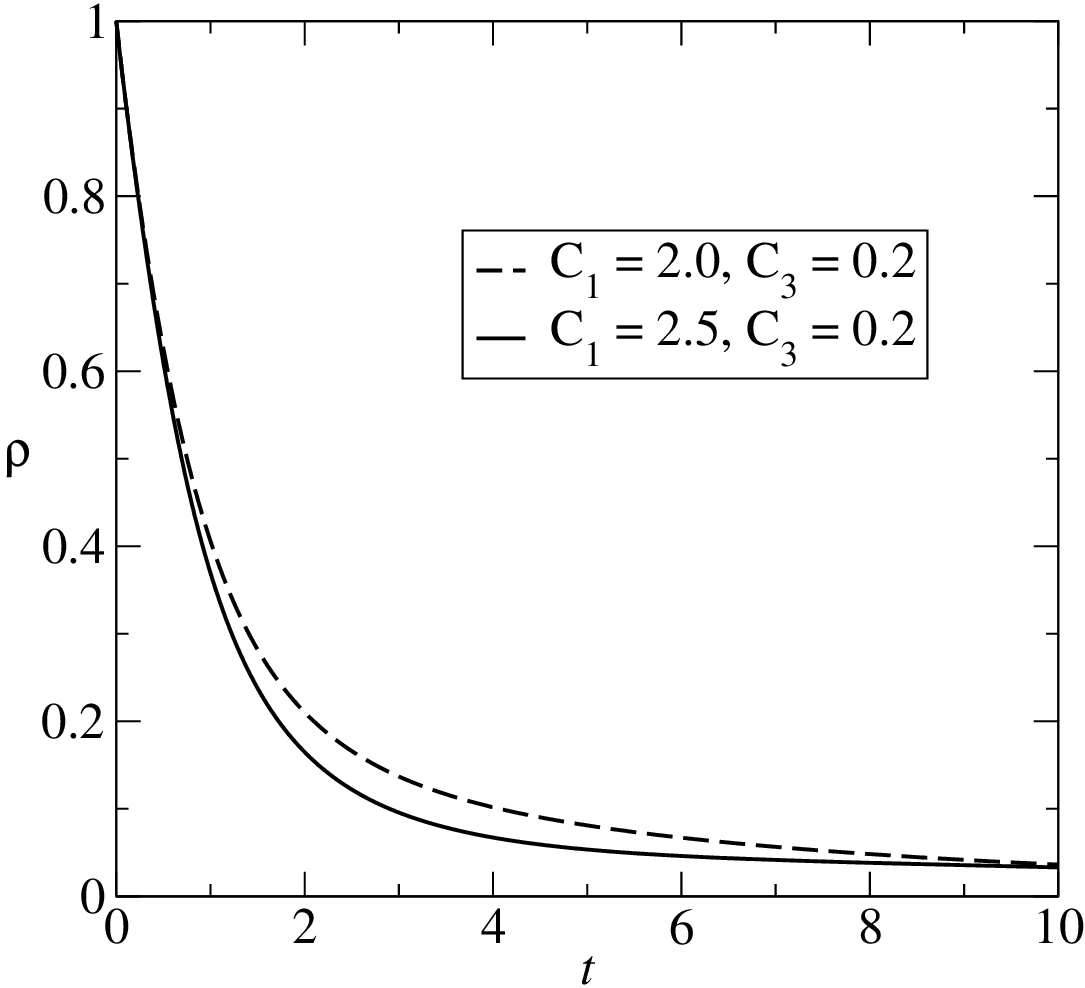}
\caption{Energy density $\rho$ versus time $t$.}
\end{center}
\end{figure}

\begin{figure}\vskip1cm
\begin{center}
\includegraphics[width=7cm]{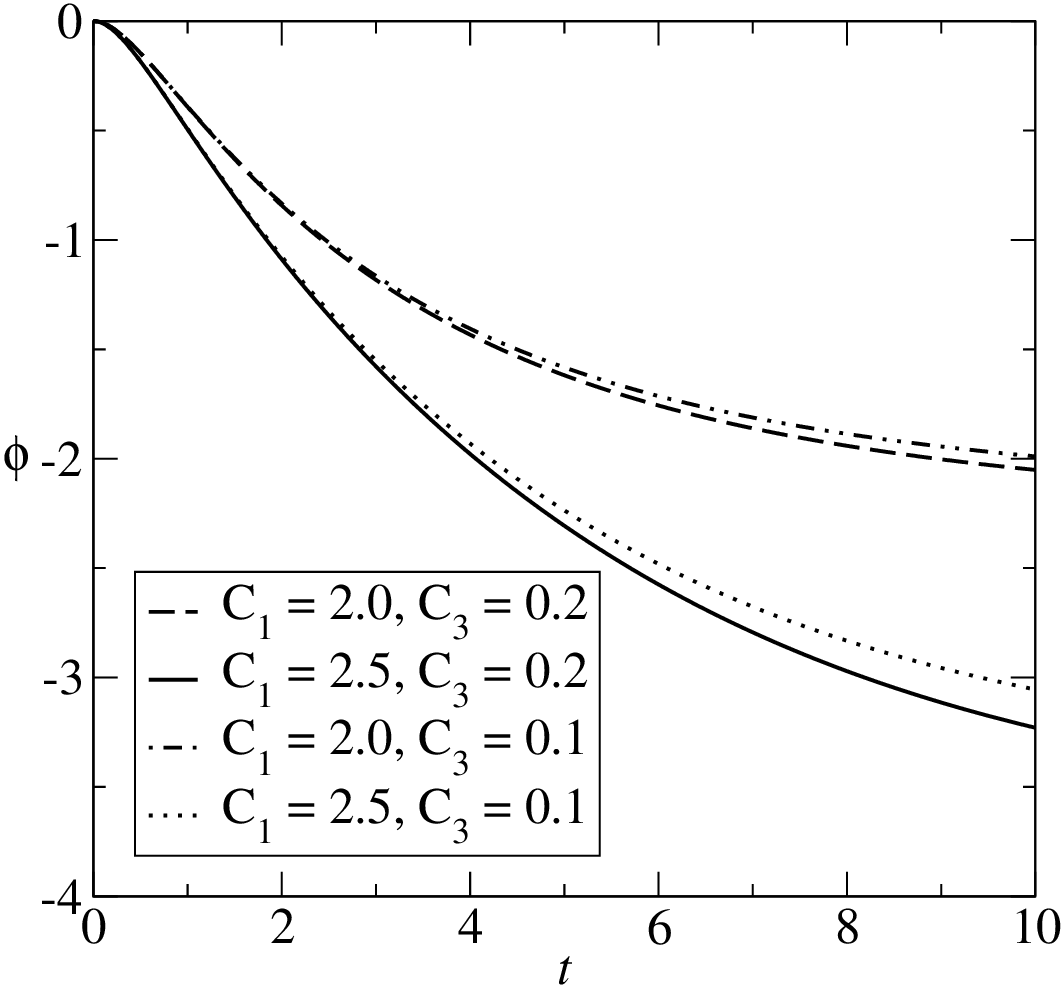}
\caption{Scalar field $\phi$ versus time $t$.}
\end{center}
\end{figure}

\section{Cosmological results}

 We have chosen for the cosmic scale factor $a(0)=1$ and for the energy density
 $\rho(0)=1.$ This set of values, combined with the Friedmann equation (\ref{14}), implies that $\dot a(0)=1/\sqrt{3}.$ Let us assume that at $t=0$ the
 scalar field  and its time variation are small, so that we choose $\phi(0)=10^{-4}$ and $\dot\phi(0)=10^{-2}$. These previous choices imply that the coupling constant $C_4$ (say) associated with the self interaction potential is determined by equation for the energy density (\ref{15}) in terms of the constant $C_1$, $C_2$ and $C_3$, namely,
 \ben\nonumber
 \frac{C_4}{a(0)^{3n}}=\rho(0)-\frac{\dot\phi(0)^2}{2}-\frac{C_1\phi(0)}{a(0)^3}-\frac{C_2\phi(0)^2}{2}-\frac{C_3}{4a(0)^6}.
 \een

There still remains to specify the free parameters, the exponent $n$ and the constants $C_1$, $C_2$ and $C_3$. The following values have been chosen:
$n=1/2$,  and $C_2=10^{-8}$, and we have investigated the role of the vector field and of the Yukawa potential in the solutions of (\ref{23}) and (\ref{24})
by varying the constants $C_1$ and $C_3$.

{
The above parameters are chosen  to represent qualitatively the transitions between accelerated and decelerated regimes, since we are working with normalized quantities (scale factor, time, energy densities). The system of equations that emerge from the original dynamics is highly non-linear, so the main point here is to identify  the
range of values  that permit the end of the inflationary period and the beginning of the matter decelerated era, taking now into account the presence of the new Walecka-type interaction.}

Numerical integrations show that the cosmic scaling factor leads to an ever expanding universe. The division of the cosmological eras can be done in terms of the acceleration field $\ddot a$. Figure 1 shows the solutions for this field according to different values of the constants: $C_1$ which is proportional to the coupling constant of the Yukawa potential and $C_3$, which is proportional to the square of the coupling constant between the fermionic and vector fields and inversely proportional to the mass of the vector field. Therefore we have  taken the values $C_1=2.0; 2.5$  and  $C_3=0.1; 0.2$. It is possible to note the existence of three cosmological eras: beginning with an accelerated expansion that can be identified with the exit of an inflationary regime, there follows a deceleration and later, a dark-energy dominated era when the universe accelerates again. It can be seen that the amplitude of the initial acceleration as well as the time length of the decelerated regime are dependent upon the choice of the constants $C_1$ and $C_3$: the greater $C_1$ or $C_3$, the smaller the amplitude of the initial acceleration and the longer the deceleration of the universe and the shorter initial acceleration. Therefore, the greater the vector field mass and the Yukawa coupling the longer the initial acceleration and the shorter the decelerated regime. It is worth noting that in the future the deceleration parameter tends asymptotically to a constant value. {The search for the values of the free parameters and initial conditions that lead to regimes where  a transition accelerated-decelerated-accelerated occurs is a very hard task, due to the instability of the non-linear coupled system of differential equations (\ref{23}) and (\ref{24}).}

The transition accelerated-decelerated-accelerated can be better understood when we plot the pressures of the fermionic, scalar and vector fields  (\ref{pp}) and the total energy density (\ref{ee1}) as functions of time. These plots are shown in Figures 2 and 3.  Whereas the amplitude and length of cosmological regimes are sensitive to different values of the constant $C_3$, the total energy density and the pressures are not, but only for different values of $C_1$. First we note that the pressure of the fermionic field is always negative, while the scalar and vector pressure fields are positive. Although the total pressure is always negative, in order to understand the  accelerated-decelerated-accelerated transition we have to analyze the acceleration field which is given by $\ddot a=-[\rho+3(p_\psi+p_\phi+p_A)]/6$. At early times the scalar and vector pressure fields are small in comparison with the fermionic pressure field so that the modulus $\vert p_\psi\vert>(\rho+3p_\phi+3p_A)/3$, which leads to a positive acceleration. At intermediate times, the density and the modulus of the fermionic pressure decay with time, while  the scalar and vector pressures increase. Hence, there exists a time interval where $\vert p_\psi\vert<(\rho+3p_\phi+3p_A)/3$ and the universe enters into a decelerated regime. After this interval the energy density and the pressures of the scalar and vector field become small and  a situation where  $\vert p_\psi\vert>(\rho+3p_\phi+3p_A)/3$ is recovered, i.e., the universe returns to an accelerated phase. {Note that due to the small value of $C_2$ the term $-C_2\phi^2/2$ does not contribute significantly to the pressure of the scalar field. The change of $p_\phi$ for different values of $C_1$ and $C_3$ is due to behavior of the term $\dot\phi^2/2$, which can be understood by observing the slopes of the graphics  $\phi\times t$ in Figure 4. Furthermore, there is no significant changes in the pressure of the vector field for different values of $C_1$ and $C_3$ chosen here, so that we have represented in Figure 2 only one curve. }

In Figure 4 it is plotted the scalar field as function of time for different choices of $C_1$ and $C_3$. The decay of $\phi$ with time is more accentuated for large values of $C_1$ and $C_3$. In all cases the scalar field tends asymptotically to a finite value for large times. {The asymptotic behavior of the scalar field $\phi$ for large values of time can be understood through the analysis of the Klein-Gordon equation (\ref{23}). For large values of time the cosmic scaling factor is also large so that the Klein-Gordon equation reduces to $\ddot\phi+3\dot a\dot\phi/a=0$, since $C_2$ was considered as a small quantity and the term $C_1/a^3$ can be neglected. Hence, the integration of the Klein-Gordon equation furnishes $\dot\phi\propto 1/a^3$, which is also a small quantity for large times. As a consequence, the pressure of the scalar field becomes very small for large values of time and the total pressure is dominated by the pressure of the fermionic field, which is negative and contributes to a positive acceleration.} It is worthwhile noting that since  $F_{\mu\nu}=0$ identically, the massive vector field $A_\nu$ is not free to propagate.

\section{Conclusions}

To sum up, we have proposed a model in which scalar, fermionic, massive vector fields and their interactions  account for the dynamics and evolution of different cosmological regimes in a homogeneous and isotropic spatially flat universe.  {By observing the behavior of the pressure of the fermionic field with respect to the scalar and vector pressure fields, we may say that it is the responsible for the two accelerated regimes, in the early and in the late periods of the Universe. As final remarks, we can consider the above investigations as a primer for future work in fermionic cosmologies, focusing on the form of the interactions and on testing these spinorial sources using other gravity theories. The future possibilities include Walecka-type interactions in scalar-tensor (Brans-Dicke) gravity, Bianchi metric fermionic scenarios   and supersymmetric inflationary regimes, which is a work in progress.}


\begin{thebibliography}{99}

\bibitem{nature} A. Riess, et al., Astronomical Journal {\bf 116}, 1009 (1998); S. Perlmutter, et al. 1999, Astrophysical Journal {\bf 517}, 565 (1999).

\bibitem{1} P.J. Peebles and B. Ratra, Astrophys. J. {\bf 325} L17 (1988); E. Copeland, M. Sami and S. Tsujikawa,
Int. J. Mod. Phys. D{\bf 15}, 1573 (2006); S. Capozziello, M. De Laurentis, S. Nojiri, S.D. Odintsov, Phys. Rev. D{\bf 79} 124007 (2009), C. Baccigalupi, A. R. Liddle and F. Perrotta Phys. Rev. D{\bf 71} 104025 (2005).


\bibitem{2} Y. N. Obukhov, Phys. Lett. A {\bf 182}, 214 (1993);
B. Saha and G. N. Shikin, Gen. Relativ. Gravit. {\bf 29}, 1099
(1997); B. Saha, Phys. Rev. D {\bf 64}, 123501 (2001); B. Saha and T.Boyadjiev, Phys. Rev. D {\bf 69},
124010 (2004); B. Saha, Phys. Rev. D {\bf74} 124030 (2006); C. Armend\'ariz-Pic\'ion and P. B. Greene, Gen. Relativ. Gravit. {\bf 35}, 1637 (2003).


\bibitem{3} M. O. Ribas, F. P. Devecchi and G. M. Kremer, Phys.
Rev. D {\bf 72} 123502 (2005); ibid, Europhys. Lett. 81, 14001 (2008); L.L. Samojeden, F. P. Devecchi and G. M. Kremer,Phys.
Rev. D {\bf 81} 027301 (2010); R. C. Souza and G. M. Kremer, Class. Quantum Grav.,{\bf 28} 125006 (2011);
 B. Saha, Phys.Rev. D {\bf 69}, 124006 (2004); Gu Y.-Q.,Int. J. Mod. Phys. A {\bf 22} 4667 (2007); Cai Y.-F.and Wang J., Class. Quantum Grav.,{\bf 25} 165014 (2008); V. V .Kassandrov, Gravit. Cosmol., {\bf 14} 53 (2008); B. Saha and M. Visinescu., Int. J. Theor. Phys., {\bf 49} 1411 (2010); J. Wang., S.-W. Cui and C. Zhang, Phys. Lett. B, {\bf 683} 110 (2010).

\bibitem{3A}C. G. Boehmer,  Phys. Rev. D {\bf77} 123535, 2008;
D. Gredat and S. Shankaranarayanan, JCAP01 008, 2010;
A. Basak and J. R. Bhatt, JCAP06 011, 2011.



\bibitem{4} M. O. Ribas, F. P. Devecchi and G. M. Kremer, Europhys. Lett. {\bf 93}, 19002 (2011).


\bibitem{birrel}  L. H. Ryder, {\it Quantum Field Theory} (Cambridge University Press, Cambridge, 1996); N.D. Birrell and P.C. Davies, {\it Quantum fields in curved space-time} (Cambridge U. press, Cambridge,  1982); J. D. Bjorken and S. D. Drell,{\it Relativistic Quantum Mechanics} (McGraw-Hill, New York, 1984); Y. Nambu , G. Jona-Lasinio, Phys. Rev. {\bf 122},(1961) 345.



\bibitem{5} C.G.Boehmer and T.Harko, Eur.Phys.J.C {\bf50}, 423 (2007);
  T. S.Koivisto and  F. M. David, JCAP, { \bf 021} 0808 (2008); G.Alexey, M. Viatcheslav and V.Victaly,JCAP,{\bf 009} 0806 (2008); L. H. Ford, Phys. Rev.D, {\bf 40 } 967 (1989); C. Armend\'ariz-P\'icon  JCAP. {\bf 0407} 007(2004).





\bibitem{8} J.D. Walecka, Ann. Phys. {\bf83}, 491 (1974); S. A. Chin and J.D. Walecka, Phys Lett.B, {\bf 52}, 24 (1974).







\end{thebibliography}
\end{document}